\documentclass[prb,twocolumn,superscriptaddress,showpacs,epsf]{revtex4}

\usepackage{graphicx}
\usepackage{latexsym}
\usepackage{amsmath}
\usepackage{amssymb}
\usepackage{amsfonts}
\usepackage{color}
\begin{document}

\title{Direct visualization of magnetic vortex pinning in superconductors}

\author{M. Menghini}
\affiliation{INPAC-Institute for Nanoscale Physics and Chemistry,
Nanoscale Superconductivity and Magnetism $\&$ Pulsed Fields
Group, K. U. Leuven Celestijnenlaan 200 D, B-3001 Leuven,
Belgium.}

\author{R. B. G. Kramer}
\affiliation{INPAC-Institute for Nanoscale Physics and Chemistry,
Nanoscale Superconductivity and Magnetism $\&$ Pulsed Fields
Group, K. U. Leuven Celestijnenlaan 200 D, B-3001 Leuven,
Belgium.}

\author{A. V. Silhanek}
\affiliation{INPAC-Institute for Nanoscale Physics and Chemistry,
Nanoscale Superconductivity and Magnetism $\&$ Pulsed Fields
Group, K. U. Leuven Celestijnenlaan 200 D, B-3001 Leuven,
Belgium.}

\author{J. Sautner}
\affiliation{Department of Electrical and Computer Engineering,
University of Illinois, Chicago, IL 60607.}

\author{V. Metlushko}
\affiliation{Department of Electrical and Computer Engineering,
University of Illinois, Chicago, IL 60607.}

\author{K. De Keyser}
\affiliation{INPAC-Institute for Nanoscale Physics and Chemistry,
Nanoscale Superconductivity and Magnetism $\&$ Pulsed Fields
Group, K. U. Leuven Celestijnenlaan 200 D, B-3001 Leuven,
Belgium.}

\author{J. Fritzsche}
\affiliation{INPAC-Institute for Nanoscale Physics and Chemistry,
Nanoscale Superconductivity and Magnetism $\&$ Pulsed Fields
Group, K. U. Leuven Celestijnenlaan 200 D, B-3001 Leuven,
Belgium.}

\author{N. Verellen}
\affiliation{INPAC-Institute for Nanoscale Physics and Chemistry,
Nanoscale Superconductivity and Magnetism $\&$ Pulsed Fields
Group, K. U. Leuven Celestijnenlaan 200 D, B-3001 Leuven,
Belgium.}

\author{V. V. Moshchalkov}
\affiliation{INPAC-Institute for Nanoscale Physics and Chemistry,
Nanoscale Superconductivity and Magnetism $\&$ Pulsed Fields
Group, K. U. Leuven Celestijnenlaan 200 D, B-3001 Leuven,
Belgium.}

\date{\today}
\begin{abstract}
We study the vortex structure in a Pb film deposited on top of a
periodic array of ferromagnetic square microrings by combining two
high resolution imaging techniques: Bitter decoration and scanning
Hall probe microscopy (SHPM). The periodicity and strength of the
magnetic pinning potential generated by the square microrings are
controlled by the magnetic history of the template. When the
square rings are in the magnetized dipolar state, known as the
onion state, the strong stray field generated at the domain walls
prevents the decoration of vortices. SHPM images show that the
stray field generated by the dipoles is much stronger than the
vortex field in agreement with the results of simulations. Real
space vortex imaging has revealed that, in the onion state,  the
corners of the square rings act as effective pinning centers for
vortices.

\end{abstract}

\pacs{74.78.-w 74.78.Fk 74.25.Dw}

\maketitle

\section{Introduction}

The most attractive hallmark of superconductivity  from the
technological stand point is the ability to carry electrical
current without losses. Unfortunately, this fragile
dissipationless state can be destroyed as a consequence of the
motion of quantum flux lines (vortices) under the action of the
Lorentz force induced by an applied electric current. During the
last decades a considerable effort has been devoted to prevent
this vortex motion by anchoring the vortices using several
methods. The vast majority of the research so far has been
focussed on studying core and electromagnetic pinning mechanisms
.\cite{tinkham,campbell,blatter,buzdin-physc-94} The former is
related to the condensation energy saved when a vortex locates at
the position of the pinning center whereas the latter originates
when the supercurrents flowing around the vortices in a distance
of the order of the penetration length, $\lambda$,
 are deformed due to the presence of the pinning sites. In general, an enhanced vortex core pinning is achieved when the size of the pinning site is similar to the
 superconducting coherence length $\xi$.\cite{civale} In contrast to that, there is no optimum size for the electromagnetic pinning as discussed in
 Refs.\,\onlinecite{shmidt,takezawa,moshchalkov-prb-98,vinokur-prb-2000}. Lithographically introduced micrometer size holes are proved to be efficient
 pinning centers for vortices in conventional superconductors\cite{pannetier,sophie} whereas nanometer columnar defects are more suitable for high $T_c$
 superconductors.\cite{civale} Further improvement of the pinning properties can be achieved if besides the type and size of individual pinning centers, also
 the geometrical distribution of pinning sites is considered.\cite{moshchalkov-prb-98,reichhardt,kemmler,villegas,misko,silhanek}

An alternative way of creating attractive potential wells for
vortices can be realized by exploiting the magnetic interaction
between vortices and nanomagnets. The pioneer work of Alden and
Livingston \cite{alden} showed that embedded magnetic particles
can indeed give rise to a clear enhancement of the critical
currents. This early report has been later on corroborated by
experiments carried out on other compounds with a better control
of the mixing of nanoparticles into the superconducting
matrix.\cite{prozorov,rizzo,wang,suleimanov,stamopoulos,stamopoulos2,kuroda,togoulev,palau2,haindl,palau,haindl2}
More recently, it has been shown that the vortex structure can be
modified due to the interaction between vortices and a non uniform
magnetic field generated at the surface of the
superconductor.\cite{erdin,carneiro} An effective modulation of
the magnetic field at the surface has been achieved with magnetic
particles deposited in a Bitter decoration
experiment\cite{bitterpinning} and by the movement of a Bloch wall
in a ferrite garnet film in contact with the
superconductor.\cite{goa-apl-2003} On the other hand, the fast
development of the lithographic techniques made it possible to
tailor, practically at will, any desired magnetic landscape at the
sample
surface.\cite{pannetier-jmmm-1993,vanbael-prb-1999,vanlook,vanbael-prl-2001,vanbael-prb-2003,martin,gillijns,lange,villegas1,fasano}
In particular, it has been shown that single
\cite{hoffman,vanbael-prb-1999,vanlook,villegas1,gillijns,villegas2,bending-prl-2007}
and multiply \cite{silhanek-physC-2007,verellen-condmat} connected
structures of magnetic material can act as efficient ON/OFF
switchable pinning potential for vortices. The properties of this
kind of pinning potential have been explored mainly through
macroscopic measurements such as electro-transport and
magnetization. Unfortunately, these integrated response techniques
conceal the ultimate details of the microscopic world of vortex
pinning.

In this work we present results on direct visualization of vortex
patterns in a superconducting Pb film deposited on top of a
periodic array of square ring shaped micromagnets. Real space
vortex arrangements are obtained in this type of hybrid system by
combining two non-invasive high resolution imaging techniques:
Bitter decoration and scanning Hall probe microscopy (SHPM). The
Bitter decoration experiments have revealed that  the strong stray
field generated by the square rings in the magnetized state acts
as a funnel for the decoration particles thus impeding the
visualization of vortices which produce a much weaker local
magnetic field. In addition, the inevitable polarization of the
evaporated magnetic clusters due to the external field gives rise
to a field-polarity selected decoration process. One way to
overcome these limitations is to image the vortices with SHPM.
Images taken with this technique at $4.2\,$K unveil the actual
location of vortices in our superconductor/ferromagnet hybrid
system both in the as-grown and the dipolar state.

Previously, vortex imaging in a superconductor/ferromagnet hybrid
shows  a preferential location of vortices at magnetic poles of
opposite polarity (compensation effect). \cite{vanbael-prl-2001}
In contrast, in our samples vortices sit at the magnetic poles
with the same polarity. This switching of the pinning potential
can be due to the different number of vortex-antivortex pairs
generated by the magnetic template, as anticipated in
Ref.\onlinecite{milosevic-2004}.

\section{Experimental details}

The samples studied in this work consist of a square array  (4
$\mu$m period) of Co square rings fabricated by e-beam lithography
and subsequent lift-off technique. The density of vortices equals
the density of square rings at a matching field of $1.3\,$Oe. The
square rings are 250 nm wide, 25 nm thick and have a lateral size
of 2 $\mu$m. In order to avoid proximity effects, an insulating
buffer layer of 10 nm thick Ge is deposited on the magnetic array.
On top of that, a 50 nm layer of Pb is evaporated.

The electrical transport properties of the sample were studied
using a Quantum Design cryostat with conventional electronics. For
these measurements the Pb film was evaporated onto a predefined
photoresist mask patterned into a transport bridge aligned with
one of the principal axes of the square lattice of the square
microrings. An ac-current of $10\,\mu$A with a frequency of
$33\,$Hz was fed to the sample with a Keithley-6221 current source
and the voltage was recorded using a SR-7265 lock-in amplifier.
The temperature stability was within $1\,$mK.

The Bitter decoration experiments were performed at $4.2\,$K after
cooling down the sample from above the critical temperature,
$T_c$, in the presence of an applied magnetic field  perpendicular
to the sample surface ({\it field cooling} experiments, FC). In
this way an homogeneous vortex distribution all over the sample is
expected. The scanning Hall probe microscopy images were obtained
using a modified Low Temperature Scanning Hall Probe Microscope
(LT-SHPM) from Nanomagnetics Instruments. A temperature sensor and
a heater were attached to the sample holder allowing a fast
stabilization and precise measurement of the sample temperature.
As in the case of the Bitter decorations, the SHPM images were
recorded  at $4.2\,$K after FC the sample. The maximum scan area
at this temperature is $\sim13\times13\,\mu$m$^2$. The Hall
voltage was measured in ac-mode with a lock-in amplifier at a
frequency of 5600 Hz and the rms value of the Hall current was
$40\,\mu$A. The images were recorded in lift-off mode with the
Hall sensor at a distance of $\sim 1\,\mu$m from the surface of
the sample.\cite{authors-shpm}

\section{Transport properties}

In order to characterize the influence of the array of Co square
rings on the superconducting properties of the Pb film we explored
the field and temperature dependence of the resistance, $R(H,T)$.
The normal-metal/supercondutor (N/SC) phase boundaries of the Pb
with the Co square microrings in the as-grown and dipolar state
are shown in Fig.\,\ref{transport} together with the results
obtained in a co-evaporated plain Pb film. The N/SC phase boundary
lines were determined by 50$\%$ of the normal state resistance.

The dipolar state, also known as the onion state\cite{Xiaobin-03},
is the remanent state after applying a $3000\,$Oe in-plane field
along one diagonal of the square rings. In this state,
head-to-head and tail-to-tail domain walls are formed in two
opposite corners of the square rings. Figures
\,\ref{field-dipole-simul}(a) and (b) show respectively, a 2D and
3D plot of the out-of-plane component of the field, $B_z$,
generated by an individual square ring in the dipolar state
obtained by micromagnetic simulation.\cite{micromagnetic}  In the
simulation we consider a Co square ring with the same dimensions
as the one used in the experiment and a saturation magnetization
 $M_s = 140\,$emu/cm$^3$.  In the 2D plot
[Fig.\,\ref{field-dipole-simul}(a)] the red/blue (light gray/dark
gray) contrast corresponds to positive and negative intensity of
the field respectively, and the arrows indicate the direction of
the in-plane magnetization. The stray field is maximum (either
with positive or negative sign) at the external part of the
corners where the magnetic moments are head-to-head  and
tail-to-tail [see Fig.\,\ref{field-dipole-simul}(b)]. In the other
two corners there is a small modulation of the field produced by a
gentle gradient of the magnetization.

The N/SC phase boundary, $H_{c2}(T)$, of the plain film is linear
(see the solid line in Fig.\,\ref{transport}) as predicted from
the Ginzburg-Landau equations $H_{c2}=
[\Phi_0/2\pi\xi^2(0)](1-T/T_c)$.\cite{tinkham} From the slope of
this boundary we estimate the coherence length for our samples
$\xi(0)=45.5\,$nm. Since the BCS coherence length for Pb is $\xi_0
\approx 83\,$nm, we can use the dirty limit expression
$\xi(0)=0.855\sqrt{\xi_0 l}$ to find the electronic mean free path
$l$.  Using  $l=34.1\,$nm and considering a
renormalization\cite{buzdin-physC-1995} of the penetration length
due to the suppression of superconductivity induced by the
magnetic array,
 we obtain $\lambda(0)\approx
40.5\,$nm. This calculation indicates that our Pb films are
type-II superconductors. This is further confirmed by the
detection of single-quanta vortices as observed by Bitter
decorations of our samples.

As a consequence of the stray fields generated by the square
microrings in the as-grown state the critical temperature,
$T_c(H=0)$, of the sample grown on top of the array of magnetic
square microrings is lower as compared to the one of the plain
film. When the square rings are magnetized in the dipolar state
the larger stray fields of the dipoles can locally destroy the
superconducting condensate inducing a further decrease of $T_c$.
In this state, commensurability effects are observed at $\pm H_1$
and $\pm 2H_1$. It is interesting to note that at $\pm 3H_1$ the
N/SC phase boundary lines of the plain film and the dipolar state
merge together. This feature was also observed in Al samples
deposited on top of Py square rings.\cite{silhanek-physC-2007}
From this result we can estimate the number of vortex-antivortex
(V-AV) pairs generated by the stray field of the square rings in
the dipolar state. Let us assume that $n_p$ V-AV pairs per unit
cell are created by the square microring array. When $H=n_pH_1$
all induced antivortices will be compensated by the vortices
generated by the applied field. In this case the resultant vortex
state consists of $n_p$ vortices
 induced by the square rings which is equivalent to apply a
field $H=n_pH_1$ in the plain film (where no V-AV pairs can be
created). From that field on, i.e. $H=\pm 3H_1$ in our particular
case, the phase boundary of the plain film and the magnetized
sample should overlap as indeed observed. This finding suggests
that the estimated number of V-AV pairs generated by the square
rings is three.\cite{authors-np}

\section{Bitter decoration results}

The transport measurements presented above show that the array of
Co square microrings acts as an effective magnetic pinning
potential for vortices in the superconductor as long as $T
\lesssim T_c$. In order to gain information about the
vortex-magnetic dipole interaction we performed Bitter decoration
experiments at $4.2\,$K. One of the main advantages of this
technique is the possibility to visualize the distribution of
vortices, with single vortex resolution, all over the surface of
the sample (of the order of mm$^2$).

{\it As-grown state}.-A FC Bitter decoration experiment carried
out at $4.2\,$K in the as-grown state at $H=0$ reveals the
presence of a disordered state with about $45\,\%$ of the square
rings in a flux-closure state [see Fig.\,\ref{deco-as-0}(a)]. This
particular magnetic state is characterized by a small out-of-plane
component of the stray field and therefore, no agglomeration  of
decoration particles is observed on top of these square rings
[Figs.\,\ref{deco-as-0}(b) and (e)].\cite{edge-deco} In the case
of decorated square rings a variety of situations appears: the
decoration particles accumulate in opposite corners of the square
rings indicating that those loops are in a dipolar state
[Figs.\,\ref{deco-as-0}(c) and (f)]\cite{authors} or neighboring
corners appear decorated in square rings that are in the so called
horse-shoe state\cite{silhanek-physC-2007}
[Figs.\,\ref{deco-as-0}(d) and (g)]. The horse-shoe state is such
that the net dipolar moment is parallel to the side of the square
ring that connects the two decorated corners. In addition, in some
cases a single spot along one side or at one corner of the square
rings is decorated. This could correspond to the presence of
magnetic vortices that generate an out-of-plane field in their
center.\cite{silhanek-apl}

{\it Magnetized state}.- Fig.\,\ref{deco-dip}(a) shows a FC
decoration image at $H=0$ after the square rings have been
magnetized by applying a uniform in-plane field of $3000\,$Oe
parallel to one diagonal of the square rings [see the white arrow
in Fig.\,\ref{deco-dip}(c)]. As expected, the magnetic dipoles
generated by the stray fields have an average magnetic moment
along this diagonal resulting in two intercalated square lattices
of positive and negative out-of-plane fields. Since the decoration
is made on the surface of the superconductor, one would expect
that the decorated poles correspond to vortices and antivortices
located at the corners of the square rings. The same result,
namely the decoration of the corners of the ring where the stray
field is maximum, is obtained when decorating directly on top of
the Co square rings without the Pb on top (not shown).

A possible way to evidence the influence of the SC layer consists
in applying a finite field in such a way that both the magnetic
poles due to the square rings and the vortices are decorated
independently. Fig.\,\ref{deco-dip}(b) shows the result of a
magnetic decoration performed at $H=H_1$. Strikingly, the image
shows that only one corner (upper right) of the square rings is
decorated. This result can be in principle attributed to the fact
that vortices generated by the applied field sit at the positions
where the out-of-plane field of the dipole is negative.
\cite{vanbael-prl-2001} Then, the decoration particles accumulate
on the corners of the square rings where the field of the dipoles
is not compensated. Under this circumstance it is expected that
for high enough fields (i.e. strictly speaking $H>n_pH_1$)
interstitial vortices would appear and agglomeration of decoration
particles should locate in between the square rings. However,
decorations performed at higher applied fields, $H=3H_1$ and
$H=10H_1$, still show that only the upper right corners of the
square rings are decorated and no interstitial vortices are
observed. A similar result is obtained in the case of a decoration
performed at $H=0.5H_1$. As an example, Fig.\,\ref{deco-dip}(c)
shows an image taken in the neighborhood of the boundary between
the patterned and non-patterned regions in a decoration experiment
made at $H=3H_1$. As expected, in the non-patterned region
vortices arrange forming a distorted triangular array with a
density corresponding to the applied field.

Notice that there are two important points in these results.
Firstly, the lack of interstitial vortices in the case $H>n_pH_1$
and secondly the decoration of a single magnetic pole for
decorations made at $H\neq0$. In order to understand the origin of
the observed lack of interstitial vortices it is necessary to
compare the field generated by the magnetic dipoles with that
associated with vortices. On the surface of the sample, the field
generated at the center of the vortex is
approximately\cite{tinkham}

\begin{equation}
2H_{c1}=\frac{\Phi_0}{2\pi\lambda^2} \ln (\lambda/\xi)
\end{equation}
where $\Phi_0$ is the flux quantum.  Using the values of $\lambda$
and $\xi$ estimated for our sample we find that the maximum field
generated by a vortex at the surface of the sample at $4.2\,$K is
about $200\,$Oe. On the other hand, from the micromagnetic
simulations we can estimate  the  stray field induced by the
square rings at the surface of the superconductor (this
corresponds to a distance of $50\,$nm above the square ring
surface). This calculation shows that the maximum out-of-plane
field (at the corner of the square ring) is $\sim 1000\,$Oe i.e.,
5 times larger than the field of a vortex. Consequently, in a
decoration experiment the evaporated magnetic particles ``feel''
the dipolar field that act as a funnel accumulating most of the
particles at the positions of the poles preventing the decoration
of interstitial vortices. This is confirmed by the decoration at
the border of the pattern shown in Fig.\,\ref{deco-dip}(c). At
first sight it seems that there is a region free of vortices right
next to the first row of Co square rings. However, a close
inspection of the image shows a row of decoration clumps in that
area  much less brighter than the rest. Vortices that are at a
distance of the order of $4\,\mu$m from the edge of the pattern
are decorated in a normal way. This indicates that one magnetic
pole can attract magnetic particles that are at a maximum distance
of $\sim6\,\mu$m (i.e. larger than the separation between magnetic
poles).

The issue that still remains unsettled is why only one pole of the
square rings is decorated when a finite magnetic field is applied.
This observation can be ascribed to the fact that the magnetic
moment of the evaporated particles aligns with the direction of
the applied field at a distance quite far away from the sample
surface as schematically represented in Fig.\,\ref{decoration-1}.
When clusters of polarized particles interact with the dipolar
field generated by the square microrings they are attracted to the
pole with parallel field orientation and repelled by the other
one. Consequently, the absence of decoration particles on the
lower left corner of the square rings is not necessarily due to
the presence of vortices but due to the broken symmetry imposed by
the applied magnetic field. This is confirmed by the result of a
decoration experiment in a sample of Co square rings without a
superconducting layer on top. In this case we perform the
decoration with the square rings in the dipolar state and a
perpendicular applied field of $10\,$Oe. One would expect that in
this situation both poles will appear equally decorated since the
applied field is not strong enough to modify the field
distribution generated by the square rings. However, it is
observed (not shown) that the decoration particles only accumulate
in the corners of the square rings where the stray field is
parallel to the applied field. {\it This finding suggests that it
is not possible to decorate either  vortices or the stray field
generated by magnetic structures when the local magnetic field
direction is opposite to the one of the applied field.}

\section{Scanning Hall probe microscopy results}

{\it As-grown state}.- A possible way to overcome this inherent
limitation of the Bitter decorations can be achieved by using a
less invasive high resolution technique: Scanning Hall probe
microscopy (SHPM). Figure\,\ref{shpm-all}(a) shows an image
obtained by SHPM after {\it field cooling} (FC) a sample with a
square array of Co square rings  in the as-grown state in a field
$H\approx1.5H_1$. In the lower left part of the image there is a
bright and a dark spot next to each other (see circles in the
image). In addition, over the right edge of the picture there is
also a dark spot. These features appear in an image taken at zero
applied field (Figure\,\ref{shpm-all}(b)) as well as in images
taken at different applied fields indicating that they correspond
to the out-of-plane stray field generated by the square
microrings. After identifying the location of the square rings
\cite{authorsrings} (dotted squares in the image) it is possible
to see that the pair of bright and dark spots is due to the stray
field of a square ring in a horse-shoe state (similar to the case
shown in Fig.\,\ref{deco-as-0}(d)). Due to the large intensity
difference between the field emanating from the magnetic poles and
that produced by the vortices, it is difficult to identify flux
lines directly in the image. Therefore, in order to determine the
location of vortices it is necessary to remove the contribution of
the field from the square rings. This is achieved by subtracting
from the raw images the field intensity picture taken at $H=0$
(see Fig.\,\ref{shpm-all}(b)) where no vortices induced by the
applied field are present. The result of this substraction is
shown in Fig.\,\ref{shpm-all}(c) where the vortices are seen as
bright spots. We identify the local maxima of intensity as the
vortex positions and depicted them as black dots in
Fig.\,\ref{shpm-all}(d). In most of the cases, vortices sit at the
corners of the square rings forming a rather disordered structure.
In the case of the square ring  in the horse-shoe state a vortex
sits along the side of the square ring in between the two poles.
It is interesting to note that the values of the magnetic
induction of the raw image spans from  -10 to 6 G [see the scale
bar of Fig.\,\ref{shpm-all}(a)] while after the subtraction of the
$H=0$ image the range of intensity is limited to the range 1 to 5
G [Fig.\,\ref{shpm-all}(c)]. This difference in scale of magnetic
field intensity is due to the stronger field induced by the square
microrings as compared to the weaker field emanating from the
vortices.

{\it Magnetized state}.- A different situation appears after
magnetizing the sample by applying an in-plane field of $3000\,$Oe
parallel to one of the diagonals of the square rings, see arrow in
Fig.\,\ref{shpm-all}(e). In this case a periodic array of magnetic
dipoles is formed as clearly seen in the image shown in
Fig.\,\ref{shpm-all}(e) obtained after FC the sample down to
$4.2\,$K in a field $H\approx1.5H_1$.

Similar to the case of the as-grown sample, a substraction of the
$H=0$ image  [Fig.\,\ref{shpm-all}(f)] is necessary in order to
identify vortex positions [see Fig.\,\ref{shpm-all}(g)]. From the
field induction scale bars of Figs.\,\ref{shpm-all}(e) and (g) we
can see that the intensity of the dipolar fields is approximately
5 times larger than the vortex field intensity, in good agreement
with the estimation made in the previous section.

In Figs.\,\ref{shpm-all}(h), (i) and (j) the vortex distributions
at $H\approx1.5H_1$, $H\approx-1.5H_1$ and $H\approx3H_1$,
respectively, are shown. The location of the centers of the
positive and negative magnetic poles are indicated by white
circles and crosses, respectively. The results of our SHPM
experiments show that in the case of FC in a positive field
$\approx1.5H_1$, most of the vortices locate on top of the
positive poles [Fig.\,\ref{shpm-all}(h)]. The fact that not all of
the positive poles are occupied could be due to the influence of
intrinsic pinning or due to small inhomogeneities in the sample.
When the sign of the field is reversed most of the vortices (with
negative polarity) sit on top of the negative poles
[Fig.\,\ref{shpm-all}(i)].

Previous SHPM studies on Pb samples deposited on top of an array
of Co/Pt {\it dots} with out-of-plane magnetization evidence that
vortices sit mainly at positions where the stray field of the
micromagnet is parallel to the vortex field.
\cite{vanbael-prb-2003} This behavior could be understood
considering that the dots generate a pinning potential for
vortices
\begin{equation}
U_m(\textbf{r})=-\int_{V_f}\textbf{m(r$^{\prime}$)}\cdot
\textbf{B(r-r$^{\prime}$)}d\textbf{r$^{\prime}$} \label{magforce}
\end{equation}
where $\textbf{m}$ is the magnetization of the dot, $\textbf{B}$
is the field of the vortex and the integration is done over the
volume of the ferromagnet. This potential has a minimum on top of
the dots when the applied field $H$ (and consequently the field of
the vortices) is parallel to the magnetization direction. On the
other hand, similar experiments in Pb samples with arrays of bars
with in-plane magnetization show \cite{vanbael-prl-2001} also
field polarity dependent pinning but now due to the interaction
between vortices and vortex-antivortex pairs generated by the
magnetic bars. In this case vortices sit at the poles where the
stray field is opposite to the applied field. Theoretical
calculations of the interaction between vortices and a ferromagnet
with in-plane magnetization demonstrate \cite{milosevic-2004} that
whether the vortex will be pinned by one or the other pole of the
micromagnet depends on the strength of the micromagnet
magnetization. In the case that the magnetization is large enough
to induce a vortex-antivortex pair in the superconductor the
equilibrium position of a vortex generated by an applied field is
the result of the balance between vortex-micromagnet,
vortex-vortex and vortex-antivortex interaction. It is important
to mention that in the calculations it is considered only the
interaction of a single vortex with one magnetic microstructure
that can generate at maximum one vortex-antivortex pair.
Unfortunately, so far there is no prediction of the lowest energy
vortex state considering also the interaction with other vortices
generated by the applied field and with more than one induced
vortex-antivortex pair. In our experiments we observe that at low
densities vortices sit mostly on top of the corners of the square
rings [Figs.\,\ref{shpm-all}(h) and (i)]. Therefore we can assure
that the magnetic square microrings act as effective pinning
centers for vortices even when lowering the temperature down to
$4.2\,$K.

Finally, when the field is further increased, SHPM FC experiments
show that vortices start to locate also at the negative poles of
the micromagnets, see for example the vortex distribution at
$H\approx3H_1$ shown in Fig.\,\ref{shpm-all}(j). Few vortices sit
at the corners of the square rings where the out-of-plane field is
very small (for example see the square microring at the upper left
corner of the image) but no vortices are seen to locate in the
interstitial sites in between the square rings. A possible
explanation for the observed vortex configuration is that vortices
placed at the positive poles generate a caging potential
\cite{caging} with a minimum in the location of the negative
magnetic poles. As a counterpart, the force exerted on the
vortices by the negative magnetic poles [see Eq.\,\ref{magforce}]
would tend to repel vortices from those positions. Thus, the
observed vortex distribution at higher densities seems to be a
result of the balance between vortex-vortex and vortex-magnetic
pole interaction. Our SHPM experiments in the magnetized state
show that the corners of the square rings, both with magnetization
parallel and opposite to the vortex field, represent effective
pinning sites for vortices.

\section{Conclusions}

We have studied the influence of an array of Co square microrings
on the superconducting properties of a Pb film. From  transport
measurements we observe that the stray fields induced by the
magnetic loops cause a reduction of $T_c$ with respect to a plain
Pb film in agreement with earlier reports. Furthermore, when the
square rings are in a dipolar state, matching features are
observed. This indicates that the micromagnets act as pinning
centers for vortices near the N/SC phase boundary line. On the
other hand, Bitter decorations at $4.2\,$K in the as-grown state
show that almost half of the square rings are in a flux-closure
state. In the dipolar state, the strong stray magnetic field
associated with the square rings act as funnels for the magnetic
decoration particles. Due to this effect, the particles accumulate
on the corners of the square rings preventing the decoration of
vortices. Our results also suggest that it is not possible to
decorate regions where the local magnetic field direction is
opposite to the one of the applied field. SHPM experiments in the
as-grown state show that, at low temperatures, the square rings
act as pinning centers even if they are in a flux-closure state.
Besides, the images obtained in FC experiments confirm that the
field induced by the square rings in the dipolar state is much
larger than the field of the vortices. This result is consistent
with the results obtained by micromagnetic simulations. Moreover,
FC images taken in the dipolar state and at low vortex densities
show that the poles with magnetization parallel to the applied
field create strong pinning sites for vortices. At higher fields,
the opposite poles can also act as preferential sites for vortices
suggesting that the resultant vortex structure is a consequence of
the competition between vortex-vortex and vortex-magnetic dipole
interaction. Our results indicate the necessity of a theoretical
model of a Superconductor/Ferromagnet hybrid system which takes
into account vortex-vortex interactions and the interaction of a
vortex with more than one induced vortex-antivortex pair.

We would like to thank F. de la Cruz, Y. Fasano, R. Luccas, P.
Mispelter and P. De Greef for useful technical discussions and W.
Gillijns for carefully reading of the manuscript. This work was
supported by the Research Fund K.U. Leuven GOA/2004/02, the Fund
for Scientific Research-Flanders FWO-Vlaanderen, the Belgian
Inter-University Attraction Poles IAP, the European ESF NES
programs, the U.S. NSF, grant ECCS-0823813 and CNM ANL grants
Nr.468 and Nr.470. A.V.S. is grateful for the support from the
FWO-Vlaanderen.

\newpage

\begin{figure}[tbb]
\caption{Normal/Superconductor phase boundary for a plain Pb film
(line) and a Pb film on top of a square array of Co square
microrings in the as-grown state (open circles) and in the dipolar
state (solid squares). The temperature is normalized by
$T_c=7.2\,$K and the field by the matching field $H_1=1.3\,$Oe.}
  \label{transport}
\end{figure}

\begin{figure}[ttb]
\caption{(a) 2D and (b) 3D plot of the out-of-plane field, $B_z$,
generated by one magnetic square ring in the dipolar state
obtained by micromagnetic simulation.\cite{micromagnetic} The
dimensions of the square ring are the same as in the experiments
and the values of the field are calculated at the surface of the
square ring. The arrows in (a) indicate the direction of the
in-plane magnetic moment.}
  \label{field-dipole-simul}
\end{figure}

\begin{figure}[htt]
\caption{(a) Scanning electron microscope (SEM) image of a Bitter
decoration performed in the as-grown state at $H=0$ and
$T=4.2\,$K. Zoom-in of panel (a) at the particular square rings
where a flux-closure (b), a dipolar (c) and a horse-shoe (d) state
are observed. (e)-(g) Schematics of a square ring in three
different states shown in (b)-(d), respectively. The yellow (light
gray) arrows indicate the average direction of the in-plane
magnetization. The red (gray) and  blue (black) dots in (f) and
(g) denote the positive and negative magnetic pole, respectively.
}
  \label{deco-as-0}
\end{figure}

\begin{figure}[htt]
\caption{Bitter decoration with Co square rings magnetized in the
dipolar state at $T=4.2\,$K and (a) $H=0$, (b) $H=H_1$ and (c)
$H=3H_1$. In the last case the image is taken at the boundary
between the patterned and non-patterned region. The white arrow
indicates the direction of the in-plane field used to magnetize
the square rings.}
  \label{deco-dip}
\end{figure}

\begin{figure}[ttt]
\caption{Schematic presentation of the accumulation of polarized
decoration particles on top of one corner of the Co square
microrings (see text). The black arrows indicate the stray field
induced by the magnetized square ring while the gray arrow denotes
the direction of the applied field. On the upper part of the image
the filament that generates the decoration particles (red spheres)
is shown.}
  \label{decoration-1}
\end{figure}

\begin{figure}[ttt]
\caption{(a) and (b) Raw SHPM images taken after FC down to
$T=4.2\,$K in a field $H\approx1.5H_1$ and $H=0$ respectively,
with the square rings in the as-grown state. The circles indicate
the positions of strong stray fields generated by the square rings
that are in a magnetized state. (c) Vortex image in the as-grown
state at $H\approx1.5H_1$. This image is obtained by subtracting
the image (b) ($H=0$) from (a). In (d) the black dots indicate the
vortex locations that were identified as local maxima in (c). (e)
and (f) Raw images taken after FC in a field $H\approx1.5H_1$ and
$H=0$ respectively, with the square rings in the dipolar state.
(g) Vortex image at $H\approx1.5H_1$ obtained by subtracting (f)
from (e). Vortex location in the dipolar state at (h)
$H\approx1.5H_1$, (i) $H\approx-1.5H_1$ and (j) $H\approx3H_1$ .
In all images the dotted lines indicate the positions of the
square rings, the field of view is $12.6 \times 12.6 \,\mu$m$^2$
and the color (gray) scale bars correspond to the magnetic
induction in units of Gauss. In (e) the black arrow indicates the
orientation of the in-plane field used to magnetize the square
microrings.}
  \label{shpm-all}
\end{figure}


\begin{references}
\bibitem{authors-np} Note that this is an approximate estimate.
\bibitem{edge-deco} The bright edges of the square microring shown in
Fig.\,\ref{deco-as-0}(b) are due to  charging effects during the
acquisition of the image with the Scanning Electron Microscope
(SEM).
\bibitem{authors}In a decoration experiment regions with positive and negative
out-of-plane local magnetic field are not distinguished from each
other since the decoration particles are sensitive to the field
gradient independently of the field direction.
\bibitem{authorsrings}In order to determine the location of the
square rings an out-of-plane field of the order of 5000 Oe was
applied after doing the FC experiments in the as-grown state.
Subsequently a SHPM image was taken where the square rings appear
as bright spots due to the finite out-of-plane component of the
magnetization.
\end{references}
\end{document}